\documentclass[a4paper,11pt]{article}
\usepackage{pos}
\usepackage{bbold}
\usepackage{tikz}
\usepackage{float}
\usepackage{booktabs}
\usepackage{multirow}

\newcommand{\dell}{\partial}
\newcommand{\hvp}{\hat{\Pi}(Q^2)}

\newcommand{\order}{\mathcal{O}}

\usepackage{braket}

\usepackage{slashed}
\newcommand{\D}{{\rm d}}
\newcommand{\Qh}{\hat{Q}}

\definecolor{darkred}{rgb}{0.8,0.1,0.1}

\usetikzlibrary{shapes,snakes,arrows,positioning,automata,backgrounds,calc,er,patterns}
\usetikzlibrary{shadings}

\makeatletter
\def\maketag@@@#1{\hbox{\m@th\normalfont\normalsize#1}}
\makeatother

\title{Hadronic contribution to the running of the electromagnetic coupling constant from lattice QCD: challenges at short distances}
\ShortTitle{Hadronic contribution to the running of $\alpha$ from lattice QCD: challenges at short distances}

\author*[f]{Sophie Mutzel}
\author[]{for the Budapest-Marseille-Wuppertal collaboration}
\affiliation[f]{Aix Marseille Univ, Universit\'e de Toulon, CNRS, CPT, IPhU, Marseille, France}

\emailAdd{mutzel@cpt.univ-mrs.fr}

\abstract{The electromagnetic coupling constant, $\alpha$, is one of the fundamental parameters of the Standard Model (SM). Its value at the Z boson mass, $\alpha(M_Z)$, is of particular interest as it enters electroweak precision tests. When running $\alpha$ from low energies up to the Z mass, five orders of magnitude in precision are lost. This makes it one of the least well determined parameters of the SM at that scale. The largest source of error comes from non-perturbative hadronic effects in the low energy region. These non-perturbative effects can be determined from ab-initio calculations in lattice QCD. At higher energies, needed to match onto QCD perturbation theory, discretization errors become large. In addition, the hadronic  vacuum polarization receives logarithmically-enhanced cutoff effects \cite{Mainzlog,Sommer2022}  which render the continuum extrapolation more difficult. To better control this extrapolation at higher energies, we test a number of improvement procedures based on lattice perturbation theory. To illustrate their effect, we present a preliminary analysis of the light quark, connected contribution to the Adler function at Euclidean $Q^2=5$~GeV$^2$. The lattice results are obtained using simulations with $2+1+1$ flavors of staggered fermions at physical values of the quark masses.}

\FullConference{The 39th International Symposium on Lattice Field Theory, LATTICE2022
8th-13th August 2022
Bonn}
\begin{document}
\maketitle

\section{Introduction}
One approach to search for physics beyond the Standard Model is at the intensity frontier,~i.e. by increasing the luminosity rather than the energy scale of experiments. Since there is still no conclusive direct or indirect evidence for new physics, we expect its effects to be very small. Detecting a deviation from the Standard Model requires both the theoretical prediction and the experimental measurement to be determined extremely precisely. On the theoretical side, hadronic effects often limit the precision, thus requiring a reliable description of these low energy QCD processes.

One of these precision quantities is the electromagnetic coupling constant, $\alpha$, whose value at the Z-mass scale is an important input in electroweak precision tests of the SM \cite{Gfitter}. While the world average of $\alpha$ at the scale of the electron mass is known to an amazing fraction of a part per billion uncertainty \cite{PDG2020}, it loses five orders of magnitude in precision when it is run up to $M_Z$ \cite{Jegerlehner}. This makes $\alpha (M_Z)$ one of the least well determined input parameters of the SM. 
Future colliders will significantly reduce the uncertainties on $\alpha(M_Z)$ obtained indirectly via fits to electroweak precision observables. In fact, the uncertainties on the direct calculation of $\alpha(M_Z)$ will have to be reduced by a factor of two to fully leverage these future measurements in the search for new fundamental physics \cite{Jegerlehner}. 
\section{The running of the electromagnetic coupling}
The full propagator of a photon with momentum $q$ is given by one-particle irreducible insertions of the self-energy tensor, denoted $\Pi_{\mu \nu}(q)$, which is the vacuum expectation value of the time ordered product of the correlator of two electromagnetic currents. By Lorentz and gauge invariance this vacuum polarization tensor can be decomposed into a Lorentz invariant function and a Lorentz structure which is transverse and proportional to two powers of $q$,
\begin{align}
\Pi_{\mu \nu}(q) &=i\int d^{4} x\ e^{i q \cdot x}\left\langle 0 | T \lbrace  J_{\mu}(x) J_{\nu}(0)\rbrace |0\right\rangle=\left(q_{\mu} q_{\nu}-g_{\mu \nu} q^{2}\right) \Pi\left(q^{2}\right) \; ,
\end{align}
with $J_{\mu}(x) \equiv \sum_{f} q_{f} \bar{f}(x) \gamma_{\mu} f(x)$, where $f = \{e, \mu, \tau, u,d,s,c,b,t\}$ with electric charges $q_f$. After resummation of the one-particle irreducible diagrams one can define the effective electromagnetic coupling:
\begin{align}
\alpha(q^{2})=\frac{\alpha(0)}{1-\Delta\alpha(q^{2})}\;, \ \ \ \ \text{with}  \ \ \
\Delta\alpha(q^{2})=4 \,\pi \, \alpha\hat{\Pi}(q^{2})=4 \,\pi \, \alpha \left(\Pi(q^{2}) - \Pi(0)\right)\;.
\end{align}
It is convenient to split the contributions to the running into a leptonic part, the hadronic contribution from the five lightest quarks and the contribution from the top quark,
 \begin{align}\label{eq:splitContributions}
\Delta\alpha(q^{2})=\Delta\alpha_{\rm{lep}}(q^{2})+\Delta\alpha_{\rm{had}}^{(5)}(q^{2})+\Delta\alpha_{\rm{top}}(q^{2}) \; .
\end{align}
While both the leptonic contribution and the contribution from the top quark can reliably be calculated in perturbation theory, the hadronic vacuum polarization function (HVP) receives large non-perturbative contributions below a scale of a few GeV, making it inaccessible to known analytic methods. Due to the complications in computing this hadronic contribution it dominates the uncertainty in the running of $\alpha$. The traditional way of obtaining the HVP in this non-perturbative regime is to use measurements of the $e^+e^- \to$ hadrons cross section as a function of the centre-of-mass energy and a dispersion relation \cite{DHMZ,KNT2019}. However, for space-like momenta, the HVP is accessible to ab-initio calculations in lattice QCD and thus provides a complementary approach which does not depend on cross-section data \cite{Borsanyi2017,BMWc2020,Mainzalpha}. This has become particularly important since the recent determination of the muon $g-2$ by the BMW collaboration, for which the predictions obtained via the data-driven approach and from the lattice differ \cite{BMWc2020}.
\section{The hadronic contribution to the running of $\alpha$ from lattice QCD}
On the lattice, the quantity that we are interested in is the Euclidean vacuum polarization tensor $Q^2=-q^2$
\begin{align}\label{eq:VPEucl}
\Pi_{\mu \nu}(Q) =\int d^{4} x \ e^{i Q \cdot x} \left\langle J_{\mu}(x) J_{\nu}(0)\right\rangle =\left(Q_{\mu} Q_{\nu}-\delta_{\mu \nu} Q^{2}\right) \Pi\left(Q^{2}\right) \; ,
\end{align}
where the last equality follows from using $\mathcal{O}(4)$ and gauge invariance. 
However, equation~\eqref{eq:VPEucl} does not hold in finite volume -- the VP is fully transverse only in infinite volume and infinite time. 
As emphasized in \cite{Bernecker}, $\hvp$ can be obtained in the time-momentum representation by a zero-momentum-projected correlator multiplied by a $Q^2$-dependent kernel function. We hence define $\forall Q{\in}\mathbb{R}$
\begin{align}\label{eq:TMR}
   \hat\Pi(Q^2)\equiv \Pi(Q^2) -\Pi(0)  =2a\sum_{t}\;\operatorname{Re}\left[\frac{e^{iQt} - 1}{Q^2}  + \frac{t^2}2\right]\,\operatorname{Re} C(t)=2a\sum_{t}k(t,Q^2)\,\operatorname{Re} C(t)  \; ,
   \end{align}
   with
   \begin{align}
   C(t) &=\frac{1}{3} \sum_{i=1}^{3} \sum_{\vec{x}}\left\langle J_{i}(x) J_{i}(0)\right\rangle =C^{u d}(t)+ C^{s}(t)+C^{c}(t)+ C^{\mathrm{disc}}(t) \; ,
\end{align}
where we have flavor decomposed the electromagnetic current correlator $J_{\mu}$.
This has the advantage that the very different statistical and systematic uncertainties of the various contributions can be addressed separately.
   
While formally Eq.~\eqref{eq:TMR} can be used to define $\hat\Pi(Q^2)$ for any $Q \in \mathbb{R}$, we are limited at large $Q$ by the momentum cutoff on the lattice, and at small $Q$ the observable will feel the finite size of the lattice. At those large distances, finite-volume and, since we are using staggered quarks, taste-breaking effects will play a role. Due to the subtraction of $\Pi(0)$ the HVP mixes, for large $Q^2$, very different scales. Hence, in this exploration of systematic effects, we instead consider the Adler function \cite{Francis_2013}, 
\begin{equation}\label{eq:Adlergeneral}
D(Q^2) \equiv 12\pi^2 Q^2\frac{\mathrm{d}\hvp}{\mathrm{d}Q^2}=24\pi^2 Q^2 \sum_{t}\frac{\mathrm{d}k(t,Q^2)}{\mathrm{d}Q^2}C(t)=\sum_{t}k_D(t,Q^2)C(t) \; .
\end{equation}
The advantage is that, for massless quarks, $Q^2$ alone determines whether the Adler function is a short- or long-distance quantity. In what follows we will focus on the challenges which show up in $D(Q^2)$ for large Euclidean $Q^2$ \footnote{Concerning the challenges and possible improvements at large distances, see for instance the supplementary material in \cite{BMWc2020}.}.
\section{Challenges at short distances: discretization errors}\label{sec:aImprovement}
At large values of $Q^2$, discretization effects become important, eventually spoiling the continuum extrapolation of the Adler function. In addition, as recently shown in \cite{Mainzlog,Sommer2022}, the contribution of the light valence quarks of mass $m_l$ to the Adler function receives logarithmically-enhanced $\mathcal{O}(a^2)$ lattice artefacts, even at leading order (LO) in lattice perturbation theory,
\begin{align}\label{eq:Symanzikexpansion}
D\left(Q^{2}, a\right)&=D\left(Q^{2}\right)\left\{1+\Gamma_0 (a Q)^{2}  \ln (a Q)^{2} + \order \left[ (aQ)^2, (am_l)^2 \right]\right\} \; ,
\end{align}
where $D\left(Q^{2}\right)$ is the value of the Adler function in the continuum and $\Gamma_0$ is a constant. These logarithmically-enhanced cutoff effects arise from small separations between the two currents; $D(Q^2)$ is not an on-shell quantity. Note also that perturbative corrections to the logarithmically-enhanced term, of the form $\alpha_s^n(1/a)(aQ)^2 \ln (aQ)^2$, are of order $\alpha_s^{n-1} (1/a) (aQ)^2$ because $\alpha_s\sim -1/\ln(a\Lambda_{\mathrm{QCD}})$ and are therefore no longer logarithmically enhanced.
\begin{figure}[t]
\begin{center}
\includegraphics[scale=0.7]{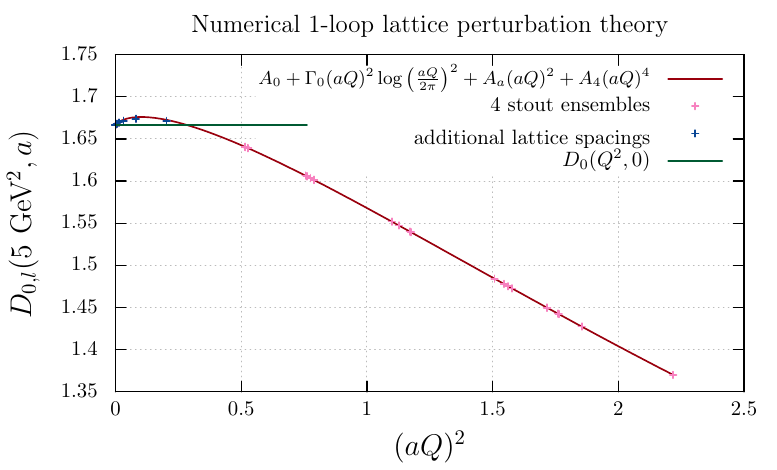}
\end{center}
\caption{Light contribution to $D(5$ GeV$^2)$ as a function of $(aQ)^2$ at LO in staggered, lattice perturbation theory. The pink points correspond to the lattice spacings available in our 4-stout ensembles \cite{BMWc2020}, blue points are additional, smaller lattice spacings. The green line is the continuum value known analytically at LO.}\label{fig:aQcont}
\end{figure}
In Figure~\ref{fig:aQcont} we plot the light contribution of the Adler function at $Q^2=5$~GeV$^2$ at LO in staggered, lattice perturbation theory as a function of the lattice spacing squared. Clearly, a naive extrapolation using a simple linear function in $(aQ)^2$ would completely miss the continuum limit. Even worse, the logarithmic term becomes important for small lattice spacings and the function turns around. Note also that this turnover is shifted towards smaller $a$ for larger values of $Q^2$. 
Hence, in order to ensure a reliable continuum extrapolation even at large $Q^2$, it is crucial to have an analytic understanding of the asymptotic dependence on the lattice spacing. For the leading-order coefficient of the logarithmically-enhanced discretization error in staggered, lattice perturbation theory we obtain
\begin{equation}\label{eq:Gamma0}
\Gamma_{0} =-\frac{1}{30}\; .
\end{equation}
\subsection{Removal of discretization effects at leading order in $\alpha_s$}
One way to tackle large cutoff effects consists in removing some of the discretization errors by improving the data using lattice perturbation theory. Hence, we define the LO improved Adler function,
\begin{align}\label{eq:Dtreeimproved}
\tilde{D}(Q^2,a)\equiv D(Q^2,a) + D_0(Q^2,0)-D_{0}(Q^2,a) \; ,
\end{align}
where $D_0(Q^2,0)$ and $D_0(Q^2,a)$ are the Adler function in the continuum and in staggered, lattice perturbation theory at LO, respectively. This should cure the data from the leading discretization errors at large momenta where perturbation theory works well. More importantly, it removes the leading, logarithmically-enhanced discretization errors, up to small $\alpha_s(Q^2)$ suppressed terms. We find for the connected zero-momentum, current correlator at LO in staggered, lattice perturbation theory
\begin{align}\label{eq:Cconndisc}
&C_0(t,a)= \frac{n_c q_f^2}{3} \sum_i \int_{-\pi/a}^{\pi/a} \frac{d\vec{p}}{(2\pi)^3} \frac{\cos(a p_i)^2 e^{-2Et}}{4\dot{E}^2}\left[ \hat{E}^2-\hat{p}_i^2 \left(1+(-1)^t\right) \right]\;,
\end{align}
with $\hat{p}_i=\sin(ap_i)/a$, $\hat{E}=\sinh(aE)/a$, $E=\operatorname{arcsinh} (a\sqrt{\sum_i \hat{p}_i^2+m^2})/a$ and $\dot{E}=\hat{E}\sqrt{1+(a\hat{E})^2}$.
This expression can be decomposed into a contribution which approaches the continuum result in the limit $a\to 0$ and a part which is oscillating around this solution (the conserved current is time-local). The disconnected component vanishes at leading order. The Adler function in infinite volume and infinite time can then be obtained from the modified Fourier transform of Eq.~\eqref{eq:Adlergeneral}
\begin{align}
D_0(Q^2,a)&= \sum_{t=0}^\infty k_D(t,Q^2) C_0(t,a)\;,
\end{align}
where the sum in $t$ can be calculated analytically. We find
\begin{equation}\label{eq:Dstagpert}
\begin{split}
&D_0(Q^2,a)= - \frac{24\pi^2}{3} n_c q_f^2\sum_i \int_{-\pi/a}^{\pi/a} \frac{d\vec{p}}{(2\pi)^3} \frac{\cos(ap_i)^2}{4\dot{E}^2} \times\\ 
&\left(\hat{E}^2\left[
\frac{\dot{E} \sin ^2\left(\frac{a Q}{2}\right)}{a \hat{E}^2 (a Q)^2 \left(-2 (a \hat{E})^2+\cos (a Q)-1\right)}+\frac{\dot{E} \sin (a Q)}{2 Q \left(-2 (a \hat{E})^2+\cos (a Q)-1\right)^2}
\right] \right.\\
&\left.-\hat{p}_i^2\left[\frac{\left(2 (a \hat{E})^2+1\right) \sin ^2(a Q)}{a \dot{E}(a Q)^2 \left(-8 (a \dot{E})^2+\cos (2 a Q)-1\right)}+\frac{4 a \dot{E} \left(2 (a \hat{E})^2+1\right) \sin (2 a Q)}{a Q \left(-8 (a \dot{E})^2+\cos (2 a Q)-1\right)^2}\right]\right) \; .
\end{split}
\end{equation}
\begin{figure}[t]
\begin{center}
\includegraphics[scale=0.7]{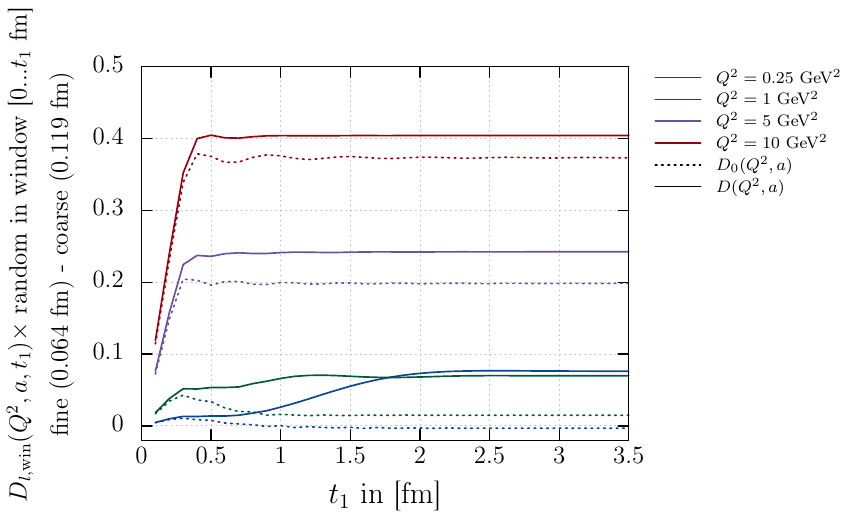}
\end{center}
\caption{Light contribution to $D(Q^2)$ in a time window $[0\ldots t_1]$~fm. We plot the difference between a fine ($a=0.064$~fm) and a coarse ($a=0.118$~fm) lattice, in comparable volumes. The solid lines are obtained from the simulation. The dotted curves are the predictions from leading-order lattice perturbation theory.}\label{fig:difftree}
\end{figure}
To investigate whether Eq.~\eqref{eq:Dstagpert} reproduces the discretization errors in our simulations for large $Q^2$, we depict in Figure~\ref{fig:difftree} the cutoff effects in the integrand of $D^l(Q^2,a)$ as a function of Euclidean time, i.e.~the integrand is convoluted by a so-called window function,
\begin{align}
D_{\rm{win}}(Q^2,a,t_1)\equiv \int \D t\ w(t ; 0, t_1) \ C(t) \ k_D(t,Q^2) \;,
\end{align}
with $w\left(t ;0, t_1\right) =\left( \tanh \left[\left(t-t_1\right) / \Delta\right]- \tanh \left[t / \Delta\right]\right)/2$ and with $\Delta=0.15$~fm. We plot this window quantity for two of our 4-stout simulations (solid line), taking the difference between the finest lattice and the coarsest lattice. These discretization errors are compared to those obtained using lattice perturbation theory for the same lattice parameters (dotted curves). For small $Q^2$, the discretization errors are small in general but they increase at large distances due to taste-breaking effects. Clearly, for times longer than 0.3~fm Eq.~\eqref{eq:Dstagpert} fails to describe the discretization errors properly. However, for large $Q^2$, discretization errors become large, they come from short distances and saturate quickly. In fact, here, LO lattice perturbation theory describes the discretization errors to better than $10\%$ for $Q^2=10$~GeV$^2$. 
\subsection{Removal of an additional discretization effect}
\begin{figure}[t]
\begin{center}
\includegraphics[scale=0.7]{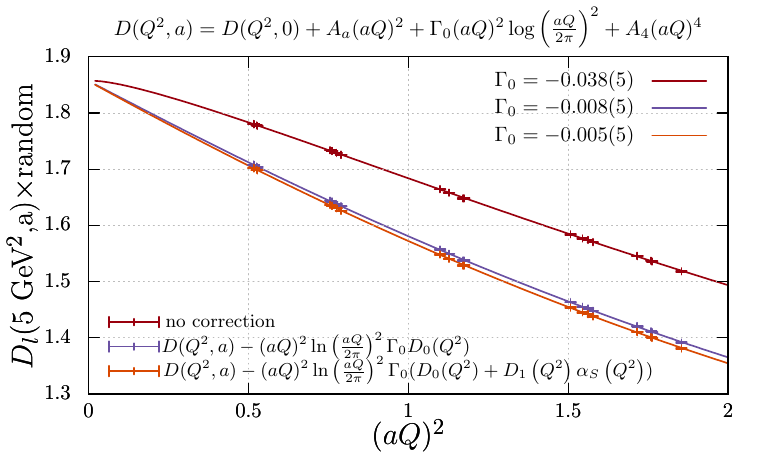}
\end{center}
\caption{Continuum extrapolation of $D_l(5$ GeV$^2)$ with the results from our 4-stout ensembles \cite{BMWc2020}. The lattice results have been blinded by a random factor between $1.01$ and $0.99$. The dark red points correspond to unimproved data, for the purple line we have removed the leading logarithmically-enhanced cutoff effect and from the orange datapoints we have removed the additional mixed term. We also depict the logarithmic coefficient $\Gamma_0$, cf.~\eqref{eq:expansionD}, obtained from a fit to the data, using the fit function in the title.}\label{fig:aQcontaddsub}
\end{figure}

Let us factorize the expansion of the Adler function as in Eq.~\eqref{eq:Symanzikexpansion}
\begin{align}\label{eq:expansionD}
&D\left(Q^{2}, a\right)=D\left(Q^{2}\right)\left\{1+\Gamma_0(a Q)^{2} \ln \left(\frac{a Q}{2 \pi}\right)^{2}+\mathcal{O}\left((aQ)^{2}\right)\right\} \; ,
\end{align}
with
\begin{align}
D\left(Q^{2}\right)=D_{0}\left(Q^{2}\right)+D_{1}\left(Q^{2}\right) \alpha_{s}\left(Q^{2}\right)+\mathcal{O}\left(\alpha_{s}^{2}\right) \; ,
\end{align} 
where $D_0(Q^2)$ and $D_1(Q^2)$ are the LO and one-loop Adler function in the continuum, respectively \cite{Chetyrkin1996}. 
Since $\Gamma_0$ and $D_1(Q^2)$ are known, by expanding equation \eqref{eq:expansionD}, we find that we can define an additionally subtracted $\tilde{D}(Q^2)$ which will have yet formally smaller logarithmically-enhanced discretization errors,
\begin{align}\label{eq:Daddimproved}
\bar{D}\left(Q^{2}, a\right) &=\tilde{D}\left(Q^{2}, a\right)-(a Q)^{2} \ln \left(\frac{a Q}{2 \pi}\right)^{2} \Gamma_{0} D_{1}\left(Q^{2}\right) \alpha_{S}\left(Q^{2}\right) \; ,
\end{align}
where $D_{0}\left(Q^{2}, a\right)$ is the leading-order improved Adler function defined in Eq.~\eqref{eq:Dtreeimproved}. After removal of this additional discretization error, logarithmically-enhanced discretization errors begin at $\order \left(\alpha_{s}^2(Q^2) (aQ)^2 \ln (a Q)^{2} \right) $ and $\order \left( (aQ)^4 \ln (aQ)^2 \right)$ and should be small. All the other terms are regular and begin at  $\order \left( (aQ)^2\right)$. In Figure~\ref{fig:aQcontaddsub} we plot the continuum extrapolation of $D_l(5$~GeV$^2)$. Two types of improvements are shown: one where the leading logarithmically-enhanced cutoff effect is removed and one where we also subtract the additional mixed term in eq.~\eqref{eq:Daddimproved}. As a fit to the unimproved lattice results shows (dark red line), the logarithmic coefficient is close to the one expected from lattice perturbation theory, see Eq.~\eqref{eq:Gamma0}. Removing the logarithmic term at leading order divides this coefficient by a factor of $\sim$five (violet line), removing the additional discretization effect reduces the logarithmic cutoff effect further: it vanishes within errors (orange line). 
\subsection{Taking the continuum limit using the lattice momentum $\Qh$}
As can be observed from Figure~\ref{fig:aQcont}, since the logarithmically-enhanced cutoff effect becomes important for small values of $(aQ)^2$, the asymptotic form of the function changes direction and approaches the continuum value from above. In order to reliably capture this turnover behaviour, results at small enough values of $(aQ)^2$ are needed. Reassuringly, as discussed in the previous paragraph, for $Q^2=5$ GeV$^2$, the logarithmic term that we obtain by a fit to the data is close to the value expected from lattice perturbation theory. One can further test that the leading logarithmically-enhanced discretization error is correctly picked up in fits to the lattice results by modifying the kernel function, as we discuss now.

A bosonic propagator on the lattice can be written in terms of a momentum $\Qh=2/a\sin \left(aQ/2\right)$. Thus, on the lattice one can define the Adler function as
\begin{align}
D\left(\Qh^2\right)&=12 \pi^2 \Qh^2 \frac{\dell \hat{\Pi}(\Qh^2)}{\dell \Qh^2}=24\pi^2 \Qh^2 \int_{0}^{\infty} \D t \ \frac{\D k(\Qh^2,t)}{\D \Qh^2} C(t)\; ,
\end{align}
with
\begin{align}
k(\Qh^2, t)=\frac{\cos (Q t) -1}{\Qh^2}+\frac{t^2}{2}\; .
\end{align}
Interestingly, the leading-order coefficient in front of the $\order (a^2)$ logarithmically-enhanced cutoff effect changes, as now also $k(\Qh^2,t)$ receives $\mathcal{O}(a^2)$ corrections. Indeed, expanding in powers of $a^2$, we find (again, neglecting mass discretization effects)
\begin{align}
\left.D_{0,f}\left(\Qh^{2},a\right)\right|_{a^{2}}=24\pi^2 \Qh^2\int_0^{\infty} \D t\left( \left.C_{0,f}(t)\right|_{a^{2}} \left.\frac{\D k(\Qh,t)}{\D \Qh^2}\right|_{a^{0}}+\left.C_{0,f}(t)\right|_{a^{0}}  \left.\frac{\D k(\Qh,t)}{\D \Qh^2}\right|_{a^{2}}\right) \; .
\end{align}
Computing the relevant integrals analytically we obtain
\begin{align}
\left.D_{0,f}\left(\Qh^{2},a\right)\right|_{a^{2}} \xrightarrow[m \to 0]{a \to 0}n_c q_f^2 \Qh^2 \ln \left(\frac{a \Qh}{2 \pi}\right)^{2} \left(-\frac{1}{12}  +\frac{1}{20} + \frac{1}{12}\right) \; ,
\end{align}
where the first discretization error is common to all staggered current correlators, the second one is specific to the conserved current and the last one originates from the kernel function.
\begin{figure}
\begin{center}
\includegraphics[scale=0.65]{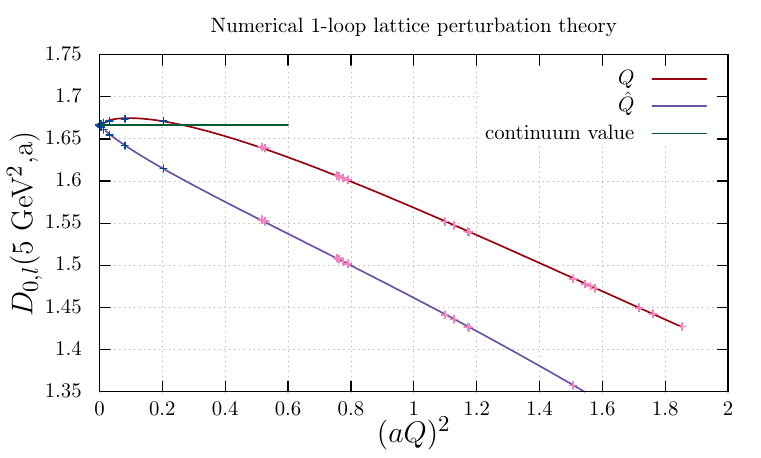}
\end{center}
\caption{Light contribution to $D_0(5$ GeV$^2)$ as a function of $(aQ)^2$ in leading-order lattice perturbation theory, employing $k(Q,t)$ (pink) and $k(\Qh,t)$ (violet) in eq.~\eqref{eq:TMR}. The pink points correspond to the lattice spacings available from our 4-stout ensembles \cite{BMWc2020}, blue points are additional smaller lattice spacings. The green line is the continuum value. By employing $k(\Qh,t)$ the continuum value is approached from below.}\label{fig:aQtreeQhat}
\end{figure}
This is what can be observed in Figure~\ref{fig:aQtreeQhat}, where $D_l(5$ GeV$^2)$ in lattice perturbation theory employing $k(Q,t)$ (pink) and $k(\Qh ,t )$ (violet) are depicted. Since the logarithmically-enhanced cutoff effect for $k(\Qh ,t)$ has a different sign, it approaches the continuum limit from below and can therefore serve as an additional systematic check. 
\section{Results}
In order to perform a controlled continuum extrapolation, we incorporate the various improvements mentioned in the previous section in our analysis. 
To obtain the physical result for $D_{l}(5$ GeV$^2)$ we perform a global fit which includes a continuum extrapolation, an interpolation to the physical point, where $X_l$ and $X_s$ parametrize the small difference in the quark masses from their physical values, and the determination of strong-isospin breaking (SIB) and QED corrections,
\begin{align}\label{eq:globalfit}
D(Q^2,a)=D(Q^2,0) +\underbrace{A(a)}_{\substack{\text{cont.} \\ \text{extrap.}}} + \underbrace{B X_{l}+C X_{s}}_{\substack{\text{interpolation to} \\ \text{physical point}}}+\underbrace{E \frac{M_{K_0}^2-M_{K_+}^2}{M_\Omega^2}+F e_v^2+G e_ve_s+H e_s^2}_{\substack{\text{determination of} \\ \mathcal{O}(\delta m, e^2)\text{ corrections}}} \; ,
\end{align}
with 
\begin{align}
X_l=\frac{M_{\pi_0}^2}{M_{\Omega}^2}-\left[\frac{M_{\pi_0}^2}{M_{\Omega}^2}\right]_*, \quad X_s=\frac{M_{K_\chi}^2}{M_{\Omega}^2}-\left[\frac{M_{K_\chi}^2}{M_{\Omega}^2}\right]_*, \quad M_{K_\chi}^2 \equiv \frac{1}{2}\left(M_{K_0}^2+M_{K_{+}}^2-M_{\pi_{+}}^2\right)
\end{align}
and
\begin{align}\label{eq:contextr}
A(a) = A_2 [a^2 \alpha_s^n(1/a)] + A_{2l}a^2\log(a^2) + A_4 [a^2 \alpha_s^n(1/a)]^2 \; , 
\end{align} 
where $A_2, A_{2l}, A_4, B,C,E,F,G,H$ are fit parameters. Note that we also distinguish valence, $e_v$, and sea, $e_s$, electric charges as in \cite{BMWc2020, Leti}.
\begin{figure}[t]
\begin{center}
\includegraphics[scale=0.6]{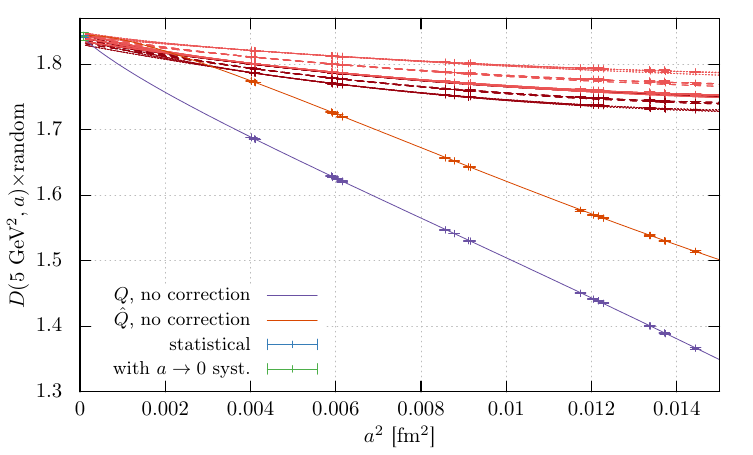}\includegraphics[scale=0.6]{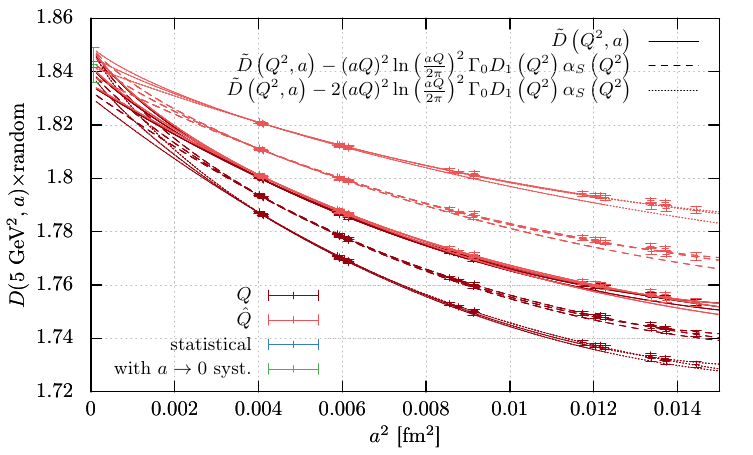}
\end{center}
\caption{Preliminary global fit of the light contribution to $D(5$ GeV$^2)$. \textit{Left:} In red the fits to the improved lattice results that enter the systematic uncertainty. In orange and violet, the fits to the unimproved lattice results employing $k(Q,t)$ and $k(\Qh,t)$, respectively. \textit{Right:} Zoom on the fits which enter the estimation of the systematic uncertainty. The linestyles indicate the various improvements. The green point at $a^2=0$ is the continuum extrapolated value with statistical and combined statistical and systematic uncertainties. }\label{fig:prelimCont}
\end{figure}
In Figure~\ref{fig:prelimCont} we show a preliminary continuum extrapolation of $D_{l}(5$ GeV$^2)$ with inclusion of SIB and QED corrections to $\mathcal{O}(\delta m, e^2)$ as detailed in \cite{BMWc2020,Leti}. It includes a systematic error originating from the continuum extrapolation, as this will be the largest source of uncertainty. The scale is set using the $\Omega$-mass, $a^2 =\left( (a M_{\Omega^-})/[M_{\Omega^-}]_*\right)^2$. We include 25 of our 26, 4-stout ensembles with lattice spacings ranging from 0.118~fm to 0.064~fm  and quark masses bracketing the physical point~\cite{BMWc2020}. To estimate the systematic error originating from the continuum extrapolation, we perform cuts in the lattice spacing, we vary the anomalous dimension (the power of the strong coupling constant in the lattice-spacing dependence in Eq.~\eqref{eq:contextr}), which is not specifically known for this quantity, from $n=0$ to $n=3$ \cite{Husung2020}. We further incorporate the following additional variations. We always improve the lattice results employing the prescription in Eq.~\eqref{eq:Dtreeimproved}, but we subtract or not the additional discretization effect of Eq.~\eqref{eq:Daddimproved} and twice this correction, we force $A_{2l}=0$ or leave it free in the fit and we employ both $k(Q,t)$ and $k(\hat{Q},t)$ as a kernel in Eq.~\eqref{eq:Adlergeneral}. The systematic uncertainty is estimated by assigning Akaike Information Criterion weights to each of the various fits. In this preliminary analysis these weights are used in a simplified fashion by taking the variance of the weighted fit values. The statistical uncertainty is estimated using the jackknife method with $N_J=48$ jackknife samples. For more details about the fitting procedure see the supplementary material in \cite{BMWc2020} (section Type-I fits). The fits to the unimproved lattice points (violet and orange) are not included in the determination of the systematic uncertainty. However, as can be observed, extrapolations using all simulations nicely lie within the estimated uncertainty. At 5~GeV$^2$, the improvement removes up to $\sim 20\%$ of the discretization effects on our coarsest lattice and allows a controlled continuum limit.
\section{Conclusion}
To determine the hadronic contribution to the running of the electromagnetic coupling using ab-initio lattice QCD calculations up to scales of a few GeV requires controlling large discretization errors and logarithmically-enhanced cutoff effects. For this purpose, we investigate a number of improvement procedures based on lattice perturbation theory. To illustrate the resulting improvements, we focus on the connected light-quark contribution to the Adler function at $5$~GeV$^2$,  for which we present a preliminary continuum extrapolation including strong and QED isospin breaking effects to $\mathcal{O}(\delta m, e^2)$. As discussed above, these improvements significantly reduce discretization errors and allow a controlled continuum extrapolation. These insights can now be applied to the short distance contributions to the running of $\alpha$. Moreover, we will perform the outlined analysis for the strange, the charm and the disconnected contribution, including a full estimation of the systematic uncertainty as in \cite{BMWc2020}. While we focused on the challenges encountered at high energies, at low momenta, finite-size effects and, since we are using staggered fermions, taste-breaking effects become important. These will be addressed in future work.
\section*{Acknowledgments}
The computations were performed on JUQUEEN, JURECA, JUWELS and QPACE at Forschungszentrum Jülich, on SuperMUC and SuperMUC-NG at Leibniz Supercomputing Centre in Munich, on Hazel Hen and HAWK at the High Performance Computing Center in Stuttgart, on Turing and Jean Zay at CNRS IDRIS, on Irène Joliot-Curie at CEA TGCC, on Marconi at CINECA. We thank the Gauss Centre for Supercomputing, PRACE and GENCI (grant 502275) for awarding time to our collaboration on these supercomputers. This research was funded in part by l’Agence Nationale de la Recherche (ANR), project ANR-22-CE31-0011, and by the Excellence Initiative of Aix-Marseille University - A*MIDEX, a French “Investissements d’Avenir” program, through grants AMX-18-ACE-005, AMX-19-IET-008 - IPhU and ANR-11-LABX-0060.

\end{document}